\documentclass[epj,twocolumn]{webofc}
\usepackage[varg]{txfonts}   % Web of Conferences font
\woctitle{NSRT15}
\newcommand{\beq}{\begin{equation}}
\newcommand{\eeq}{\end{equation}}
\newcommand{\bea}{\begin{eqnarray}}
\newcommand{\eea}{\end{eqnarray}}
\newcommand{\eps}{\varepsilon}

\begin{document}

\title{Self-consistent Theory of Finite Fermi Systems
 vs     Skyrme--Hartree--Fock method. Spherical nuclei.}

\author{E.\,E. Saperstein\inst{1}, \inst{2} \fnsep\thanks{\email{saper43\_7@mail.ru}}
\and   S.\,V. Tolokonnikov\inst{1}, \inst{3}}

\institute{National Research Centre ``Kurchatov Institute'', 123182, Moscow, Russia \and National
Research Nuclear University MEPhI, 115409 Moscow, Russia \and Moscow Institute of Physics and
Technology, 141700, Dolgoprudny, Moscow Region, Russia}

\abstract{Recent results of the Fayans energy density functional (EDF)  for spherical nuclei are
reviewed. A comparison is made with predictions of several Skyrme EDFs. The charge radii and
characteristics of the first $2^+$ excitations in semi-magic nuclei are briefly discussed. The
single-particle spectra of doubly magic nuclei are considered in more detail. The phonon-particle
coupling effects are analyzed including the tadpole term.}

\maketitle

\section{Introduction}
\vskip -0.3cm
 In this century,
 the  Hartree--Fock (HF)  method with Skyrme forces \cite{HF-VB} dominates in the
theoretical description of ground states of heavy nuclei. This method is often considered as a version
of the Energy Density Functional (EDF) method by Kohn and Sham \cite{K-Sh}, which is based on the
Hohenberg--Kohn theorem \cite{Hoh-K}, stating that the ground state energy $E_0$ of a Fermi  system ia
a functional of its density $\rho$. The Skyrme HF ansatz leads to a rather simple form of the nuclear
EDF. The same is true for the HF method with Gogny force \cite{Gogny} which becomes again popular last
decade. For completeness, we mention also the relativistic mean field (RMF) model, see recent Ref.
\cite{RMF-drip} and links there, and a  newly developed self-consistent approach, known as a method
BCPM (Barcelona - Catania - Paris - Madrid) \cite{BCPM1,BCPM6}.

We use an alternative EDF method developed by Fayans with coauthors \cite{Fay1,Fay4,Fay5,Fay} with
more sophisticated density dependence. The main, in-volume term of the Fayans EDF can be symbolically
written as \vskip -0.2cm\beq { \cal E}(\rho) = \frac{a \rho^2} 2 \frac{1+ \alpha \rho^{\sigma}}{1+
\gamma \rho^{\sigma}}, \label{Fay0} \eeq \vskip -0.2cm
\noindent where $\rho({\bf r})$ is the total
nuclear density, $a, \alpha, \sigma, \gamma$ being parameters. The corresponding term of the Skyrme
EDF corresponds to $\gamma{=}0$ in this relation. The use of the bare mass $m^*{=}m$ is another
peculiarity of the Fayans EDF. Both the features of the Fayans method are closely related to the
self-consistent Theory of Finite Fermi Systems (TFFS) and represent, in a hidden form, energy
dependence effects inherent to this approach. The latter is based on the general principles of the
TFFS \cite{AB1} with inclusion of the TFFS self-consistency relations \cite{Fay-Khod}.  The final
version of this approach \cite{Sap-Kh,Khod-Sap} was formulated in terms of the quasiparticle
Lagrangian ${\cal L}_q$, which is constructed to produce the quasiparticle mass operator
$\Sigma_q({\bf r},k^2;\eps)$. By definition, the latter coincides  with the exact mass operator
$\Sigma({\bf r},k^2;\eps)$ at the Fermi surface. In the mixed coordinate-momentum representation it
depends linearly on the momentum squared $k^2$ and the energy $\eps$ as well \cite{AB1}.

In the TFFS, the effective mass  is a product $m^*{=}m^*_k \cdot m^*_E$ of the ``k-mass'' and the
``E-mass''. The two effects compensate each other almost exactly \cite{Khod-Sap} resulting in $m^*
\simeq 1$ which justifies the Fayans choice of the bare mass instead of the effective one. The EDF of
the self-consistent TFFS is found from the Lagrangian ${\cal L}_q$ according the canonical rules. It
includes implicitly the $Z$-factor \vskip -0.2cm \beq Z({\bf r})= \left(1- \left(\frac {\partial
\Sigma} {\partial \eps}\right)_0\right)^{-1}, \label{Z-fac}\eeq \vskip -0.2cm \noindent where the
index 0 means that the energy and momentum  variables are taken at the Fermi surface. Its  density
dependence can be found explicitly \cite{Khod-Sap}: \vskip -0.2cm \beq Z({\bf r})=2\left(1+
\sqrt{1-4C_0 \lambda_{02}\rho({\bf r})/\eps_{\rm F}^0}\right)^{-1} \label{Z0},\eeq \vskip -0.2cm
\noindent where $C_0=(dn/d\eps_{\rm F}^0)^{-1}= \pi^2/m p_{\rm F}^0 $ is the usual TFFS normalization
factor, inverse density of states at the Fermi surface, and the dimensionless parameter $\lambda_{02}$
determines the $Z$-factor of nuclear matter $Z_0$. In the result, the density dependence of the EDF
becomes rather complicated \cite{Khod-Sap}. Fayans with coauthors found \cite{Fay1} that, in a wide
density region, it can be approximated  with high accuracy by a more simple expression (\ref{Fay0}).
Thus, the Fayans EDF method can be interpreted as a version of the self-consistent TFFS.

Till this year, Ref. \cite{Fay-def}, all self-consistent calculations with Fayans functionals were
carried out for spherical nuclei only. It proved out successful in systematic description of nuclear
magnetic \cite{mu1,mu2}  and quadrupole \cite{BE2,QEPJ,QEPJ-Web}  moments and nuclear radii
\cite{Sap-Tolk} as well. In the latter case, agreement with the data is better than that in all SHF
calculations we know. There are two examples more where we may compare the two approaches directly.
The energies and $B(E2)$ values for the first $2^+$ excitations in semi-magic nuclei
\cite{BE2,BE2-Web} is one of them. They    are described with the Fayans EDF much better than in the
known SHF calculations with SkM* and SLy4 EDFs \cite{BE2-HFB}.
  The single-particle energies (SPEs) $\eps_{\lambda}$   in seven doubly magic nuclei, for which the
experimental spectra are known \cite{exp},  is another example \cite{Levels}. The PC corrections to
$\varepsilon_{\lambda}$ were found self-consistently with account for the tadpole diagram. A
systematic comparison to the SHF predictions with a popular HFB-17 EDF \cite{HFB-17} was carried out.
Even at the level of the mean field theory, the Fayans EDF results are significantly better. Inclusion
of the PC corrections makes the agreement better yet.

The article is organized as follows. Sect. 2 contains a brief comparison with several Skyrme EDFs of predictions for the charge radii
and characteristics of the first $2^+$ excitations in semi-magic nuclei. Sect. 3 is devoted to the description of SPEs in doubly magic nuclei. Sect. 4 contains conclusions.

\vskip -0.4cm
\section{Charge radii and characteristics of $2^+_1$ levels in semi-magic nuclei}
\vskip -0.2cm
 In Ref. \cite{Sap-Tolk},  systematic self-consistent calculations of charge radii
$R_{\rm ch}$ were made on the base of the DF3-a EDF. Deformed nuclei were also included into analysis,
with an approximate taking into account the deformation effect. Agreement with the data on the level
of 0.01 fm was achieved, noticeably better than for SLy4 and HFB-17 Skyrme EDFs taken for comparison.
The HFB-17 predictions were taken from  \cite{site}, whereas for the SLy4 EDF calculations were made
in \cite{Sap-Tolk}. As an example, the lead charge radii are displayed in Fig. 1. We see that the
Fayans EDF, indeed, describes the radii perfectly well. The HFB-17 one reproduces the data reasonably
for heavy Pb isotopes but fails systematically for $A{<}190$, the disagreement reaching 0.1 fm. The
reason for this is that this EDF erroneously predicts rather strong stable deformation for the light
Pb isotopes which leads to a significant increase of $R_{\rm ch}$ values. This problem is discussed in
detail in \cite{Fay-def}. The SLy4 $R_{\rm ch}$ values are systematically higher than the experimental
ones at approximately 0.03 fm. This is a typical scale of accuracy in describing nuclear radii for
different Skyrme EDFs.

\begin{figure}
\resizebox{0.8\columnwidth}{!}{\includegraphics {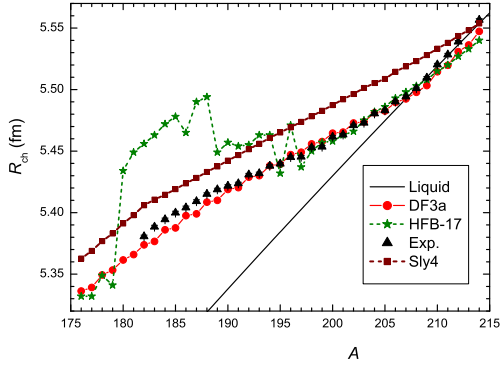}} \vskip -0.2 cm \caption{ Charge radii in lead
isotopes. Solid line shows the Liquid Drop model predictions.}
\end{figure}

\begin{figure}
\resizebox{0.8\columnwidth}{!}{\includegraphics {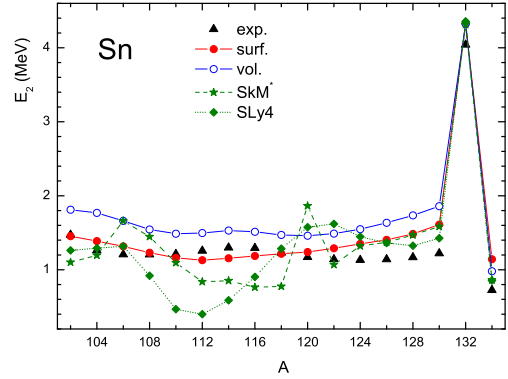}} \vskip -0.2 cm \caption{ Excitation energies
$\omega(2^+_1)$ for tin isotopes. Predictions for the SkM* and SLy4 force are taken from
\cite{BE2-HFB}.  }
\end{figure}

A high accuracy of reproducing the charge radii may be linked to more adequate density dependence of
the Fayans EDF compared to  the Skyrme   one. Indeed, if we denote the average error in describing the
binding energies as $\overline{\delta E}$ and that for the charge radii as $\overline{\delta R_{\rm
ch}}$, these quantities should be, due to the Hohenberg--Kohn theorem \cite{Hoh-K}, proportional to
each other, \vskip -0.3cm \beq \overline{\delta R_{\rm ch}}=\alpha \; \overline{\delta E}
\label{HK}\,, \eeq \vskip -0.2cm \noindent where the coefficient $\alpha$ depends on the functional we
use. This almost obvious relation can be proved with such simple consideration. Let we have an exact
EDF ${\cal E}_0(\rho)$ and add to it a small addendum $\delta {\cal E}_0=\lambda f(\rho)$, where the
coefficient is small, $\lambda\ll 1$. Obviously, we get $\overline{\delta E}=\lambda \int d{\bf r}
f(\rho({\bf r)})=\lambda a$; then the change of the mean field $\delta U=\delta f / \delta \rho$ is
proportional to $\lambda$. Let us find the change of the density $\rho$ in the first order of the
perturbation theory $\delta U$. It can be easily seen that $\delta \rho$ and the corresponding change
of the radius $\overline{\delta R_{\rm ch}}$ are also proportional to $\lambda$, $\overline{\delta
R_{\rm ch}}=\lambda b$. In the result, we obtain (\ref{HK}) with $\alpha=b/a$. As a rule, a fine
tuning of the EDF parameters is performing  by focusing mainly on reproduction of the nuclear masses
within a minimal value of $\overline{\delta E}$. In this case, the accuracy of reproducing the charge
radii is proportional to the coefficient $\alpha$. As the analysis of \cite{Sap-Tolk} showed, for the
Fayans EDF this coefficient is less than those of the HFB-17 and SLy4 functionals. This observation
may be linked to more sophisticated density dependence of Fayans functional, which allows to
incorporate implicitly the energy dependence effects.

\vskip 0.2cm
\begin{figure}
\resizebox{0.8\columnwidth}{!}{\includegraphics {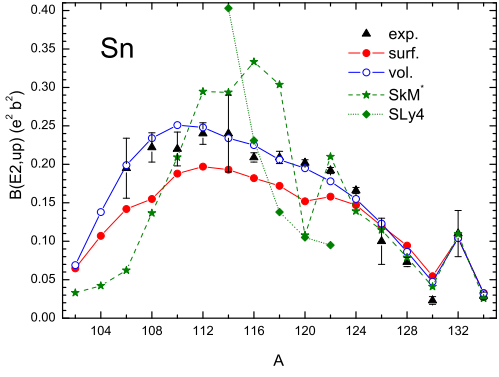}} \caption{$B(E2,{\rm up})$ values for tin
isotopes. Predictions for the SkM* and SLy4 force are taken from \cite{BE2-HFB}.   }
\end{figure}

 In Ref. \cite{BE2}, excitation energies $\omega_2$  and transition probabilities $B(E2)$ of the first $2^+$
excitations in even tin and lead isotopes as well as  the quadrupole moments  of odd neighbors of
these isotopes were calculated within the self-consistent TFFS with the use of the DF3-a EDF.  The
effect of the density dependence of the effective pairing interaction was analyzed  by comparing
results obtained by solving the QRPA equations with  volume and surface pairing. The effect was found
to be noticeable giving evidences in favor of the surface pairing. For example, the $2^+$-energies are
systematically higher at 200-400 keV for the volume paring as compared with the surface  pairing case,
the latter being on average better. Obtained results were compared with predictions of \cite{BE2-HFB}
for the Skyrme EDFs SkM* and SLy4. For tin isotopes, the comparison of $\omega_2$ and $B(E2)$ values
is given in Figs. 2 and 3, correspondingly.    On the average, predictions of the Fayans EDF with both
models for pairing reasonably agrees with the data perfectly well, significantly better than those for
both the Skyrme EDFs. Among the latter, the SkM* EDF turns out to be more successful. For SLy4 EDF,
the $^{112}$Sn nucleus and its neighbors are very close to the point of the quadrupole instability, in
contradiction with the data.

\vskip -0.3cm
\section{Single-particle spectra of magic nuclei}
\vskip -0.3cm
 The major part of modern calculations with the Fayans functional in spherical nuclei are
made with the EDF DF3-a. This is a version \cite{Tol-Sap} of the EDF DF3  \cite{Fay4,Fay} with
modified spin-dependent parameters, the spin-orbit $\kappa,\kappa'$ and the first spin harmonics
$g_1,g_1'$ which play the role of the effective tensor forces. The DF3 and DF3-a EDFs both contain 3
non-zero spin-dependent parameters, specifically the value of $g_1{=}0$ is putted. The bulk of the
data \cite{exp} for SPEs contains 65 spin-orbit differences which permits to try to find an optimal
set of the spin-dependent parameters of the  EDF. Such a set DF3-b, with $g_1{\neq}0$,  was found in
\cite{Levels}, but the overall description of these differences turned out only a bit better than for
DF3 or DF3-a sets. Moreover, the average accuracy in reproducing SPEs $\eps_{\lambda}$, see Table 1,
for DF3-b EDF is the same as for DF3 and only a little better than for the DF3-a EDF. The latter was
chosen for systematic calculations as far as it previously proved to be successful in description of
different nuclear phenomena \cite{BE2,QEPJ,Tol-Sap}. For comparison, we found the SPEs of all nuclei
under consideration for the Skyrme EDF HFB-17 \cite{HFB-17} which is a record-holder in reproducing
nuclear masses.

\begin{table}
\caption{Average deviations  $\langle\delta \eps_{\lambda}\rangle_{\rm rms}$ (MeV) of
the theory predictions for the single-particle energies from the
experimental values for magic nuclei. }
\begin{center}
\begin{tabular}{cccccc}
\hline
\hline
Nucleus   & $N$ &DF3-b & DF3-a &DF3& HFB17\\

\hline
$^{40}$Ca & 14& 1.08  & 1.25 & 1.35  & 1.64 \\

$^{48}$Ca & 12& 0.89   &1.00 & 1.01  & 1.70 \\

$^{56}$Ni & 14& 1.00   &0.97 & 0.85  & 1.40 \\

$^{78}$Ni & 11& 1.24   & 1.41 & 1.09  & 1.32 \\

$^{100}$Sn& 13& 1.09 & 1.17 & 1.01 & 1.56 \\

$^{132}$Sn& 17& 0.58  & 0.66 & 0.55 & 1.15  \\

$^{208}$Pb& 24& 0.44  & 0.51 &0.43 & 1.15 \\

\noalign{\smallskip}\hline\noalign{\smallskip}

Total & 105& 0.89  & 0.98 &0.89 & 1.40 \\
\hline
\hline
\end{tabular}
\end{center}
\label{tab:error}
\end{table}

Figs. 4 and 5 show the comparison to the data \cite{exp} of predictions of all the EDFs under
discussion for the SPEs in $^{208}$Pb for neutrons and protons, correspondingly. We see that both
HFB-17 spectra, especially the neutron one, are too expanded. This is the result of influence of the
effective mass $m^*<m$. The analysis in Ref. \cite{Dobaczewski} confirmed  a preference of the choice
of the bare mass for describing SPEs with Skyrme EDFs without PC corrections.

\begin{figure}
\resizebox{0.8\columnwidth}{!}{\includegraphics {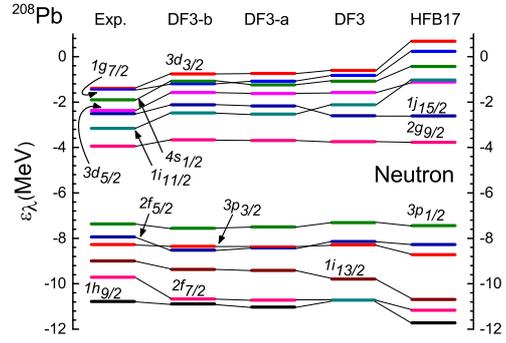}} \vskip 0.2cm  \caption{ Neutron
single-particle levels in $^{208}$Pb. Experimental data from \cite{exp}.}
\end{figure}

\begin{figure}
\resizebox{0.8\columnwidth}{!}{\includegraphics {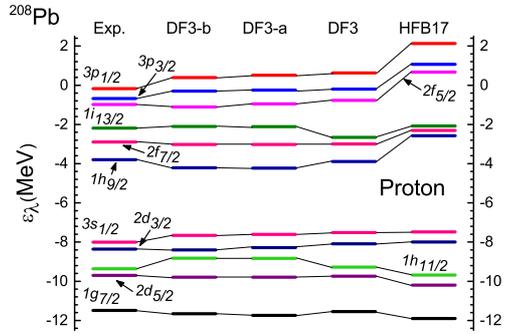}} \vskip 0.2cm \caption{ Proton single-particle
levels in $^{208}$Pb. Experimental data from \cite{exp}.}
\end{figure}

\begin{figure}[tbp]
\resizebox{0.8\columnwidth}{!} {\includegraphics  {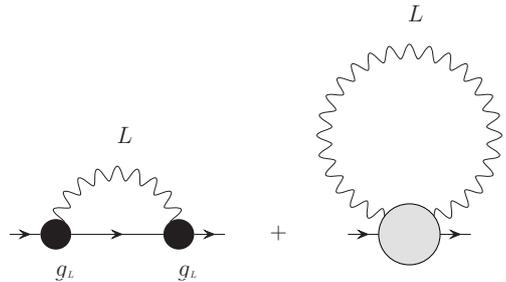}}
 \caption{PC corrections to the mass operator. The gray
circle denotes the ``tadpole'' term.}
\end{figure}

Let us go to the PC contributions to   SPEs. Accounting for PC effects, the equation for SPEs
 and corresponding wave functions  can be written as \vskip -0.3cm \beq \left(\eps-H_0
-\delta \Sigma^{\rm PC}(\eps) \right) \phi =0, \label{sp-eq}\eeq  where $H_0$ is the quasiparticle
Hamiltonian with the spectrum $\eps_{\lambda}^{(0)}$ and $\delta \Sigma^{\rm PC}$ is the PC correction
to the quasiparticle mass operator. After expanding this term in the vicinity of
$\eps=\eps_{\lambda}^{(0)}$ one finds \vskip -0.5 cm \beq \eps_{\lambda}=\eps_{\lambda}^{(0)} +
Z_{\lambda}^{\rm PC} \delta \Sigma^{\rm PC}_{\lambda\lambda}(\eps_{\lambda}^{(0)}) ,\label{eps-PC}\eeq
with obvious notation. Here $Z^{\rm PC}$ denotes the $Z$-factor due to the PC effects, i.e. that found
from Eq. (\ref{Z-fac}) with substitution of $\delta \Sigma^{\rm PC}(\eps)$ instead of the main mass
operator $\Sigma(\eps)$. Remember that in the TFFS the corresponding $Z$-factor is included in the
quasiparticle Hamiltonian $H_0$.

The PC correction to the mass operator is displayed in Fig. 6, where $g_L$ is the vertex for creating
the $L$-phonon. In magic nuclei, it obeys the equation \cite{AB1} \vskip -0.3cm \beq {
g_L}(\omega)={{\cal F}} {A}(\omega) { g_L}(\omega), \label{g_L} \eeq
 where $ A(\omega)=\int
G \left(\eps + \omega/ 2 \right) G \left(\eps - \omega/ 2 \right)d
\eps/(2 \pi i)$ is the particle-hole propagator, $G(\eps)$ being the
one-particle Green function. In obvious symbolic notation, the pole
diagram corresponds to  $\delta\Sigma^{\rm pole}=(g_L,D_L G g_L)$, where
$D_L(\omega)$ is the phonon $D$-function.

\begin{figure}[tbp]
\resizebox{0.8\columnwidth}{!} {\includegraphics  {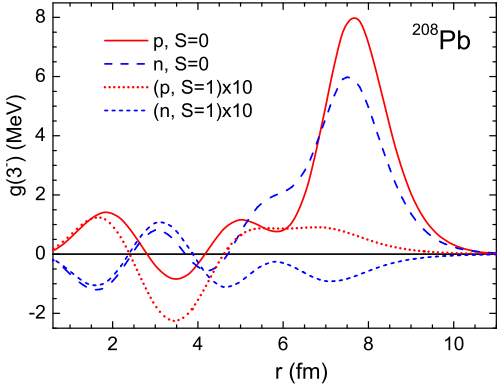}}
 \caption{The vertex $g_L$ for the
$3^-_1$ state in $^{208}$Pb.}
\end{figure}

All the low-lying phonons we consider have  natural parity. In this case, the vertex $g_L$ possesses
even $T$-parity. It is a sum of two components with spins $S=0$ and $S=1$, respectively, \vskip -0.3cm
 \beq g_L= g_{L0}(r) T_{LL0}({\bf n,\alpha}) +  g_{L1}(r) T_{LL1}({\bf n,\alpha}), \label{gLS01} \eeq
where $T_{JLS}$ stand for the usual spin-angular tensor operators. The operators $T_{LL0}$ and
$T_{LL1}$ have  opposite $T$-parities, hence the spin component should be the odd function of the
excitation energy, $g_{L1}\propto \omega_L$. For the ghost dipole, $L=1$ and $\omega_1=0$, Eq.
(\ref{g_L}), due to the TFFS self-consistency relation \cite{Fay-Khod}, has the exact solution \beq
g_1 ({\bf r}) = \alpha_1  (d U(r)/d r) Y_{1M}({\bf n}), \label{g1}\eeq where $\alpha_1=
 1 /\sqrt{2\omega
B_1}$ , $B_1=3m A/4\pi $ is the Bohr--Mottelson (BM) mass coefficient \cite{BM2} and
$U(r)$ is the central part of the mean-field potential generated
by the energy functional.

The second, tadpole, term in Fig. 6  is \beq \delta\Sigma^{\rm tad}=\int \frac {d\omega} {2\pi i}
\delta_L {g_L} D_L(\omega),\label{tad} \eeq where $\delta_L {g_L}$  can be found \cite{Khod-Sap} by
variation of Eq. (\ref{g_L}) in the field of the $L$-phonon: \bea  \delta_L {g_L}&=&\delta_L {\cal F}
A(\omega_L){ g_L} + {\cal F} \delta_
L A(\omega_L){ g_L} \nonumber \\
&+& {\cal F} A(\omega_L)\delta_L{g_L}. \label{dgL} \eea

The quantity $\delta_L A$ can be readily obtained by variation of
each Green function in the particle-hole propagator $A$ in field
$g_L$ induced by the $L$-phonon. The explicit expression for the
variation $\delta_L {\cal F}$ can not be found within the TFFS as in
this approach the Landau--Migdal amplitude ${\cal F}$ is introduced as a
phenomenological quantity. In \cite{Khod-Sap} the ansatz was
proposed, \beq  \delta_L {\cal F} = \frac {\delta {\cal F}(\rho)}
{\delta \rho} \delta \rho_L,\label{dLF}\eeq where \beq \delta \rho_L
=A_L g_L   \label{rhoL} \eeq is the transition density for
excitation of the $L$-phonon. The phonon
$D$-function appears in Eq. (\ref{tad}) after connecting  two wavy
phonon ends in Eq. (\ref{dgL}). This corresponds to averaging of the
product of two boson (phonon) operators $B_L^+B_L$ over the ground
state of the nucleus with no phonons.

The $L$-phonon excitation energies $\omega_L$ and creation amplitudes $g_L({\bf r})$ were found by
solving Eq. (\ref{g_L}) in a self-consistent way with the DF3-a functional. In more detail, the
procedure is described in \cite{BE2}.  All the $L$-phonons we consider are the surface vibrations
which belong to  the Goldstone mode corresponding to the spontaneous breaking of the translation
symmetry in nuclei \cite{Khod-Sap}. The coordinate form of their creation amplitudes $g_L({\bf r})$ is
very close to that, Eq. (\ref{g1}) , for the ghost $1^-$ phonon, which is the lowest energy member of
this mode: \beq g_L(r)=\alpha_L (dU/dr) +\chi_L(r), \label{gLonr}\eeq where the in-volume correction
$\chi_L(r)$ is rather small.

\begin{table}
\caption{Pole and tadpole contributions  (MeV) to PC corrections from
$3^-$-states to  neutron SPEs in $^{208}$Pb. }
\begin{center}
\begin{tabular}{cccc}
\hline
\hline

$\lambda$ &  $\delta \eps^{\rm pole}_{\lambda}$ & $\delta \eps^{\rm
tad}_{\lambda}$ & $\delta \eps_{\lambda}$ \\
\hline

$3d_{3/2}$ & -0.150 &  0.012 & -0.137 \\
$2g_{7/2}$ & -0.142 &  0.061 & -0.081 \\
$4s_{1/2}$ & -0.134 &  0.016 & -0.118 \\
$3d_{5/2}$ & -0.147 &  0.023 & -0.124 \\
$1j_{15/2}$& -0.708 &  0.204 & -0.504 \\
$1i_{11/2}$& -0.058 &  0.198 &  0.140 \\
$2g_{9/2}$ & -0.244 &  0.076 & -0.167 \\
$3p_{1/2}$ & -0.220 &  0.053 & -0.167 \\
$2f_{5/2}$ & -0.186 &  0.094 & -0.092 \\
$3p_{3/2}$ & -0.205 &  0.056 & -0.149 \\
$1i_{13/2}$&  0.057 &  0.211 &  0.269 \\
$2f_{7/2}$ &  0.724 &  0.091 &  0.815 \\
$1h_{9/2}$ & -0.014 &  0.197 &  0.184 \\
\hline
\hline
\end{tabular}
\end{center}
\end{table}

\begin{table}
\caption{Pole and tadpole contributions (MeV) to PC corrections from
$3^-$-states to proton SPEs in $^{208}$Pb. }
\begin{center}
\begin{tabular}{cccc}
\hline
\hline

$\lambda$ &  $\delta \eps^{\rm pole}_{\lambda}$ & $\delta \eps^{\rm
tad}_{\lambda}$ & $\delta \eps_{\lambda}$ \\
\hline

$3p_{1/2}$ & -0.375 &  0.153 & -0.222 \\
$3p_{3/2}$ & -0.371 &  0.152 & -0.219 \\
$2f_{5/2}$ & -0.278 &  0.168 & -0.110 \\
$1i_{13/2}$& -0.534 &  0.266 & -0.268 \\
$2f_{7/2}$ & -0.409 &  0.168 & -0.240 \\
$1h_{9/2}$ & -0.054 &  0.222 &  0.168 \\
$3s_{1/2}$ & -0.310 &  0.143 & -0.167 \\
$2d_{3/2}$ & -0.241 &  0.146 & -0.095 \\
$1h_{11/2}$& -0.017 &  0.246 &  0.229 \\
$2d_{5/2}$ &  0.435 &  0.147 &  0.582 \\
$1g_{7/2}$ & -0.271 &  0.197 & -0.074 \\
 \hline
 \hline
\end{tabular}
\end{center}
\end{table}

The smallness of the in-volume component $\chi_L$ is demonstrated in
Fig. 7 for the $3^-_1$ state in $^{208}$Pb, which is the
most collective one among the surface vibrations and plays the main role in PC corrections for this nucleus.
The small spin components $S=1$ are also displayed.  To make them
distinguishable, they are multiplied by the factor of 10.  The
smallness of the spin components is typical for $L$-phonons with
a high collectivity.  The first,
surface term on the right-hand sight of Eq. (\ref{gLonr}) corresponds to the
BM model for the surface vibrations \cite{BM2}, the amplitude
$\alpha_L$ being related to the dimensionless BM amplitude $\beta_L$
as follows: $\alpha_L=R \beta_L$, where $R=r_0 A^{1/3}$ is the
nucleus radius, and $r_0=1.2\;$fm.

If one neglects in-volume contributions, the tadpole PC term (\ref{tad}) can be reduced to a very
simple form: \vskip -0.1cm \beq \delta\Sigma^{\rm tad}_L = \frac {\alpha_L ^2} 2 \frac {2L+1} 3
\triangle U(r). \label{tad-L}\eeq The general consideration of the tadpole term within the not
self-consistent TFFS, with solving Eq. (\ref{dgL}), was carried out in \cite{Platon}. It was found
that  the in-volume corrections to Eq. (\ref{tad-L}) are, indeed, small for heavy nuclei, e.g., for
$^{208}$Pb. At the same time, for light nuclei, e.g., $^{40,48}$Ca, the accurate solution of Eq.
(\ref{dgL})  diminishes the approximate value (\ref{tad-L})  for the tadpole term by $\simeq 30$\%.

Following to \cite{Levels}, we neglect the in-volume corrections for all nuclei
considered. To find the phonon amplitudes $\alpha_L$,
 we use  the definition $\alpha_L^{\tau}=  {g_L^{\tau,{\rm max}}}/
   ( {dU}/{dr})^{\tau,{\rm max}},$ with obvious notation.
It should be noted that the values of $\alpha_L^n$ and $\alpha_L^p$
are always very close to each other and to that which follows from the
BM model formula for $B(EL)$: $B(EL)_{\rm BM}= \left(
3Z/4\pi\right)^2\beta_L^2 R^{2L}$ \cite{BM2}, where the
dimensionless BM phonon creation amplitude $\beta_L$ is related to that
used by us as $\alpha_L=\beta_L R /\sqrt{2L+1}$, $R=1.2\,A^{1/3}$.
For example, for the $3^-_1$ state in $^{208}$Pb we have:
$\alpha_L^n=0.32\;$fm, $\alpha_L^p=0.33\;$fm, and $\alpha_L^{\rm
BM}=0.30\;$fm.

A comparison of the pole and tadpole PC corrections to neutron and proton SPEs, induced by the $3^-_1$
state in $^{208}$, are given in tables 2 and 3, correspondingly. The tadpole term is always positive,
whereas the pole one is, as a rule, negative, and the two contributions are of the opposite sign. The
magnitude of the tadpole term is, as a rule, less than the pole one, but comparable with the latter.
Especially it this true for protons. Therefore the sum is often essentially less than the pole term
alone. A typical suppression of the pole contribution is of 30--50\%, but there are cases of a
stronger suppression, e.g. $2d_{3/2}$ and $1g{7/2}$ proton states. Moreover, there are several cases
than the tadpole term dominates: the neutron $1i_{11/2}$ and $1h_{9/2}$ states and the proton
$1h_{9/2}$ and $1h_{11/2}$ ones. In these cases, the total correction is of the opposite sign as
compared with the pole term. The general conclusion is that one overestimates the PC correction to
SPEs neglecting the tadpole term, and often it is better to omit it completely than consider the pole
term alone.

\begin{table}
\caption{PC effect on average deviations  $\langle\delta \eps_{\lambda}\rangle_{\rm rms}$ (MeV) of
the theory predictions for SPEs from the
experimental values for the DF3-a functional.}
\begin{center}
\begin{tabular}{cccc}
\noalign{\smallskip}\hline\noalign{\smallskip}

Nucleus   & $N$ & DF3-a+PC  & DF3-a \\
\noalign{\smallskip}\hline\noalign{\smallskip}

$^{40}$Ca & 14 & 1.30 &  1.25  \\

$^{48}$Ca & 12 & 1.05 & 1.00  \\

$^{56}$Ni & 14 & 0.98 &0.97  \\

$^{78}$Ni & 11 & 1.34 & 1.41 \\

$^{100}$Sn& 13 & 1.21 & 1.17  \\

$^{132}$Sn& 17 & 0.63 & 0.66   \\

$^{208}$Pb& 24 & 0.38 & 0.51  \\

\noalign{\smallskip}\hline\noalign{\smallskip}

total & 105 & 0.97 & 0.98  \\

\noalign{\smallskip}\hline\noalign{\smallskip}
\end{tabular}
\end{center} \label{tab:rmsPC}
\end{table}

The total PC effect to SPEs of all doubly magic nuclei under consideration is shown in table 4. We see
that it makes the agreement essentially better for $^{208}$Pb but a bit less good in the light nuclei.
The latter is explained mainly with the use the approximation (\ref{tad-L}) for the tadpole term.
According to \cite{Platon},  the exact consideration for $^{40,48}$Ca diminishes the tadpole term by
approximately 30\%. Such a correction should lead to a better agreement with the experimental SPEs.
Let us stress once more a high accuracy of reproducing the experimental SPEs in $^{208}$Pb:
$\langle\delta \eps_{\lambda}\rangle_{\rm rms}{=}0.38$ (MeV). Indeed, the corresponding result
\cite{Litv-Ring}, obtained within the RMF approach, is $\langle\delta
\varepsilon_{\lambda}\rangle_{\rm rms}{=}0.85$ MeV.

\vskip -0.3cm
\section{Conclusions}
\vskip -0.3cm
 Recent studies with the Fayans EDF for spherical nuclei are reviewed and compared to
predictions of several Skyrme EDFs. The charge radii and the characteristics of the first $2^+$
excitations in semi-magic nuclei are briefly discussed. The accuracy of reproducing the charge radii
is of the order of 0.01 fm, which is significantly better than that obtained with  HFB-17 and SLy4
EDFs. The excitation energies and $B(E2)$ values of the first $2^+$ states in the lead and tin
isotopic chains also agree with the data much better than the results \cite{BE2-HFB} obtained with the
SLy4 and SKM* EDFs.

The SPEs of doubly magic nuclei are considered in more detail. The PC effects in SPEs are analyzed particularly. In addition to the usual pole diagram, the consideration  includes the so-called tadpole term which is usually ignored \cite{Litv-Ring,Bort}. The latter is considered approximately, with neglecting the in-volume components of the vertex $g_L(r)$ of creating a surface $L$-phonon. This approximation  works well in heavy nuclei, leading to a simple formula for the tadpole contribution which can be easily included into the calculation scheme. The tadpole contribution  is often comparable with that of the pole diagram. As a rule, these two contributions to SPEs have different signs, their sum being often significantly less than that of the pole term alone. For the $^{208}$Pb nucleus, the calculation without PC corrections results in  the
average deviation   $\langle\delta \varepsilon_{\lambda}\rangle_{\rm rms}$  from the experimental values equal to 0.51 MeV for the Fayans EDF, whereas it is 1.15 MeV for the HFB-17 EDF.  With the PC corrections, including the tadpole contribution, we obtained very high accuracy: $\langle\delta \varepsilon_{\lambda}\rangle_{\rm rms}{=}0.34$ MeV. For a comparison, the corresponding quantity,  found in \cite{Litv-Ring} within the RMF approach with PC corrections without the tadpole term, is $\langle\delta \varepsilon_{\lambda}\rangle_{\rm rms}{=}0.85$ MeV.

To conclude, the Fayans EDF leads to a better agreement with the experimental data in all phenomena in spherical nuclei considered  than several popular Skyrme EDFs chosen for a comparison. We relate it to the peculiarities of the Fayans EDF which reflect, in a hidden form, the energy dependence effects inherent to the self-consistent TFFS.

\vskip -0.3cm
\section{Acknowledgment}
\vskip -0.3cm
 The work was partially supported  by the Grant NSh-932.2014.2 of the Russian Ministry
for Science and Education and  by the RFBR Grants  13-02-00085-a, 13-02-12106\_ofi-m, 14-02-00107-a,
14-22-03040\_ofi-m.

\vskip -25mm
\end{document}